\documentclass[12pts]{article}
\usepackage{times}
\usepackage[dvips]{epsfig}
\usepackage{paralist}
\usepackage[latin1]{inputenc}
\usepackage{psfrag,color}
\begin{document}%

\title{The Formation Mass of a Binary System via Fragmentation of a
Rotating Parent Core with Increasing Total Mass}

\author{G. Arreaga-Garc\'{\i}a\footnote{Corresponding
author:garreaga@cifus.uson.mx}\\
Departamento de Investigaci\'on en F\'{\i}sica de la Universidad de Sonora \\
Apdo. Postal 14740, Hermosillo, 83000 Sonora, M\'exico.\\}

\maketitle%
\abstract{Recent VLA and CARMA observations have shown proto-stars
in binaries with unprecedented resolution. Specifically, the
proto-stellar masses of systems such as CB230 IRS1 and L1165-SMM1 have
been detected in the range of $0.1-0.25 \, M_{\odot}$. These are
much more massive than the masses generally obtained by numerical
simulations of binary formation, around $0.01 \, M_{\odot}$. Motivated by
these discrepancies in mass, in this paper we study the formation mass of a
binary system as a function of the total mass of its parent core.
To achieve this objective, we present a set of numerical
simulations of the gravitational collapse of a uniform and rotating
core, in which azimuthal symmetric mass seeds are initially
implemented in order to favor the formation of a dense filament, out
of which a binary system may be formed by direct fragmentation. We
first observed that this binary formation process is diminished when
the total mass of the parent core $M_0$ is increased; then we
increased the level of the ratio of kinetic energy to the
gravitational energy, denoted by $\beta$, initially supplied to the
rotating core, in order to achieve the desired direct fragmentation
of the filament. Next, we found that the mass of the binary
fragment increases with the mass of the parent core, as expected. In this
paper we confirm this expectation and also
measure the binary mass $M_f$ obtained from an initial $M_0$. We
then show a schematic diagram $M_0$ vs $\beta$, where the desired
binary configurations are located while the ratio of the thermal energy to the
gravitational energy $\alpha$ is kept fixed. We also report some
basic physical data of the proto-stellar fragments that form the
binary system, including the formation mass $M_f$, and its
corresponding $\alpha_f$ and $\beta_f$. Finally, we show the
resulting velocity distribution for our calculated models.}
{\it keywords:--stars: formation;--physical processes: gravitational collapse, 
hydrodynamics;--methods: numerical}

\section{Introduction}
\label{intro}

The formation of low mass-star binaries is well-understood with regard to its
basic physical principles; see \cite{boden} and \cite{stahler}. The
essential events of this formation process are the gravitational collapse
of cores and their fragmentation during an early evolution
stage of the collapsing cores; see \cite{rei}, \cite{duch}
and \cite{gir}.

As for the observational aspect, recent technical
improvements have made it possible to improve the spatial resolution
and observe a few protostars in their Class 0 evolution stage; for
instance, L1157-mm, CB230 IRS1 and L1165-SMM1, all isolated and
located in the Cepheus Flare region; see \cite{obs}; there was less
evidence of direct observation of a disk, for instance, that
with radius 125 AU surrounding the protostar L1527; see
\cite{tobin2012}. However, spatial resolutions better than 50 AU are
needed in order to produce basic physical information and provide
clues regarding the mechanisms of binary system formation.

For the Taurus dark cloud, a correlation
between the mass of the newly formed stars and the mass of the
associated dense proto-stellar cores in the cloud was
observed long ago by ~\cite{myers}.

With regard to the theoretical aspect, numerical simulations aimed
at reproducing the collapse of rotating cores began to be performed
four decades ago. The most well-known example of isothermal
fragmentation during a core collapse was first calculated by
\cite{boss1979}. This model is now called the ''standard isothermal
test case'' as it has been used for testing new codes and making
code comparisons. The outcome of this classic model and of a variant
of thereof calculated by \cite{burkertboden93} and \cite{bateburkert97}
was a protostellar binary system.  

The earliest papers on collapse
were largely done with insufficient spatial resolution; see for
instance \cite{boss1991}, therefore, these calculations suffered from
artificial fragmentation due to violation of the Jeans condition;
see \cite{truelove}. Nowadays, a new generation of three-dimensional collapse
calculations  have improved so much in resolution, that they now reveal, in
very valuable detail, the formation of a binary system or multiple systems
of low mass proto-stars by starting with a density perturbation with
azimuthal symmetry that was successfully implemented long
ago by \cite{boss2000};  see \cite{truelove98}, \cite{klein99},
\cite{boss2000}, \cite{kitsionas} and \cite{springel}. Many of these
numerical experiments were done using a solar mass parent core, that
collapsed under its self-gravity against its thermal pressure and
rotational support to form a binary system composed of very low mass
proto-stars; see the review by \cite{tohline}.  However, taking
advantage of scaling relations valid in a nearly homologous
isothermal collapse, \cite{sterzik} demonstrated that the final
properties of the binary system depend on the
initial conditions of the parent core, such as its temperature, mass
and angular momentum. The assumption of isothermality allows for
the existence of such scaling relations, so that the results obtained
for the collapse of a one solar mass parent core can be scaled to cores
of arbitrary mass only in the isothermal regime.

Recent VLA and CARMA observations have shown proto-stars
in binaries with unprecedented resolution. Specifically, the
proto-stellar masses of systems such as CB230 IRS1 and L1165-SMM1 have
been detected in the range of $0.1-0.25 \, M_{\odot}$. These are
much more massive than the masses generally obtained by numerical
simulations of binary formation, with an initial fragment 
mass of around $0.01 \, M_{\odot}$, when the calculations must be stopped because 
of insufficient spatial resolution and small time steps. The fragments will 
accrete mass and continue to grow so long as infalling gas is available. Motivated by
these discrepancies in mass, in this paper we study the formation mass of a
binary system as a function of the total mass of its parent core.
To achieve this objective, in this paper we present high-resolution three-dimensional
hydrodynamical simulations, done with the public code
Gadget2, which implements the SPH technique, in order to follow the
gravitational collapse of a uniform and rotating core in a variant of
the standard test case, in which we
implemented a mass perturbation, with the same mathematical structure
as the density perturbation used by \cite{boss2000}, to enforce the
formation of two antipode embryonic
binary mass seeds during the early core
collapse. This system evolved to a pair of well-defined mass
condensations connected by a dense filament.

In this intermediate evolution stage of the core collapse, two events may take
place according to the assembled mass of these mass
condensations. If the assembled mass is low enough
for the centrifugal force (due to core rotation) to overcome the
gravitational attraction of the mass condensations, then they
approach each other, achieve rotational speed, swing past
each other and finally separate to form the
desired fragments, which will become the binary system,
in which the fragments orbit around one another. When the
condensed masses are massive enough, then they approach each other, make
contact and merge to form a single central proto-stellar mass condensation.

When the total mass of the parent core $M_0$ is increased, the
merging event is thus favored while the binary formation process is
diminished. The formed single central mass condensation is
surrounded by a disk out of which two additional small mass
condensations may be formed by disk fragmentation, so that this kind
of simulation ends with a multiple system, dominated by a primary
mass. These kind of configurations were obtained by
\cite{hennebelle} by increasing the external pressure on a rotating
core.  To prevent the occurrence of merging, we increased the ratio
of rotational energy to gravitational energy, $\beta$, supplied
initially to the parent core, while we kept the ratio of thermal
energy to potential energy, $\alpha$, fixed for our all simulations.
\cite{matsumoto} and \cite{tsuribe2} studied the effects
of different rotation speeds and rotation laws on the fragmentation
of a rotating core. These authors introduced six types of
fragmentation seen as the possible outcomes of a collapsing core.
The configuration that interests us in this paper
corresponds to their disk-bar type fragmentation. They also showed a
configuration diagram whose axes were given by products of the free
fall time $t_{ff}$ measured from the central density, multiplied by the
initial central angular velocity $\Omega_0$, and the amplitude of
the velocity perturbation of the $m=2$ mode $\Omega_2$,
respectively.

To start our study, we arbitrarily chose two initial $\beta$ ratios: the
lower value given by $\beta=0.045$ and the higher value given by $\beta=0.14$,
while we fixed $\alpha=0.24$. The number of Jeans masses contained in
a uniform spherical core is given by $1/\alpha$. Thus the number of
fragments that we expect to form is 2. We thus constructed a
schematic diagram the axis of which were $M_0$ and the
measured $\beta$, where we showed the kind of configuration
obtained: either primary or the desired binary. It is important to note that
recent observations made by \cite{tokovin} seem to confirm the existence of 
dwarf binaries with 
mass ratios $q=M_2/M_1$ from 0.95 to 1, as first described by \cite{lucy}; some 
examples of this kind of 
system are MM Her and EZ Peg, with $q$ given by 0.98 and 0.99, respectively. The 
particular mechanism 
of fragmentation considered in this paper can be potentially useful as a template 
for studying binary systems with $q$ close to 1.

Furthermore, as the parameter space relevant to the core collapse is
so large, it is not easy to anticipate the outcome of a given
collapse simulation. A first effort for establishing a criterion
of the type $\alpha \, \beta <  0.2 $, for predicting the
occurrence of fragmentation of a rotating isothermal core was
obtained by means of numerical simulations by \cite{miyama},
\cite{hachisu1} and \cite{hachisu2}. Furthermore, \cite{tsuribe1}
made a semi-analytical study to construct a configuration space whose
axes were the dimensionless quantities $\alpha$ and $\beta$. \cite{tsuribe2} introduced 
another fragmentation criterion based on
the flatness of the core, such that the configuration diagrams
$\alpha$ versus $\beta$ were improved. It must be emphasized that
numerical simulations seem to prove that these fragmentation
criteria can only provide a clue regarding the fate of a specific initial
core configuration but can-not predict its exact outcome or
the number of fragments that may be produced during its
gravitational collapse.

According to these fragmentation criteria, the values for $\alpha$ and $\beta$ 
that we use in this paper favor the
collapse of the core and the formation of the embryonic binary system
according to \cite{tsuribe1}, but it is not still clear what the next
events are, as they depend on the assembled mass, as we mentioned in a
previous paragraph. So, we performed numerical simulations in order
to determine exactly the main simulation outcome beyond the formation
of the mass condensations.

In the early simulations on the collapse of rotating cores, the ideal equation
of state was used as a first approximation. However, once that gravity
produces a substantial contraction of the core, the gas begins to heat. In
order to take this heating into account, in our simulations we implemented
a barotropic equation of state {\it beos}, as was
proposed by~\cite{boss2000}. The {\it beos} depends on a
single free parameter, the critical density $\rho_{crit}$.
\cite{NuestroRMAA} reported a study of the effects of the change in the
thermodynamic regime on the outcome of a particle-based simulation, where
several values of the critical density were considered. The simulations that
we present here increased the peak density up to three orders of magnitude
within this adiabatic regime. Therefore, the scaling relations
obtained under the assumption of isothermality are no
longer valid for this final evolutionary stage of our simulations.

The outline of this paper is as follows: the basic physics of the
core and the particle distribution that represents the initial core
are described in Section \ref{sec:coreinicial}. The most important
features of the time evolution of these simulations are presented by
means of 2D iso-density plots in Section \ref{sec:resultados}. The
relevance of these results in view of those reported in previous
works is discussed in Section \ref{sec:dis}. Furthermore, we show
the velocity distribution obtained for the binaries by means of
iso-velocity 3D plots in Section\ref{subsec:distri}. Finally, some
concluding remarks are made in Section \ref{sec:conclu}.
\section{The core}
\label{sec:coreinicial}

We consider a spherical core with radius
$R_0=4.99 \, \times 10^{16}$ cm $\equiv \, 3335 \,$ AU, which
is rigidly rotating around the $z$ axis with an angular velocity $\Omega$, so
that the initial velocity
of the $i-th$ SPH particle is given by $\vec{v}_i= \vec{\Omega} \times \vec{r}_i
\equiv (-\Omega\, y_i,\Omega\, x_i,0)$.

In Section~\ref{sec:resultados} we will present the results of several
simulation models where the total core mass is systematically
increased up to $5 \, \, M_{\odot}$. We emphasize that we left the
initial core radius unchanged, so that the average density increased
instead. However, these models still correspond to a core,
as defined statistically by \cite{bergin}.

The time needed for a test particle to reach
the center of the core when gravity is the only force acting on it, is defined
as the free fall time $t_{ff}$ by means of

\begin{equation}
t_{ff} = \sqrt{ \frac{3\, \pi}{32 \, G \, \rho_0}}
\label{tff}
\end{equation}
\noindent where $\rho_0 = 3.8\, \times 10^{-18}\, $ g cm$^{-3}$ is the
average density of a $1 \, M_{\odot}$ core with radius $R_0$. These values
of $t_{ff}$ and $\rho_0$ are used as normalizing factors for the plots
presented below.

\begin{figure}
\begin{center}
\includegraphics[width=4.2in]{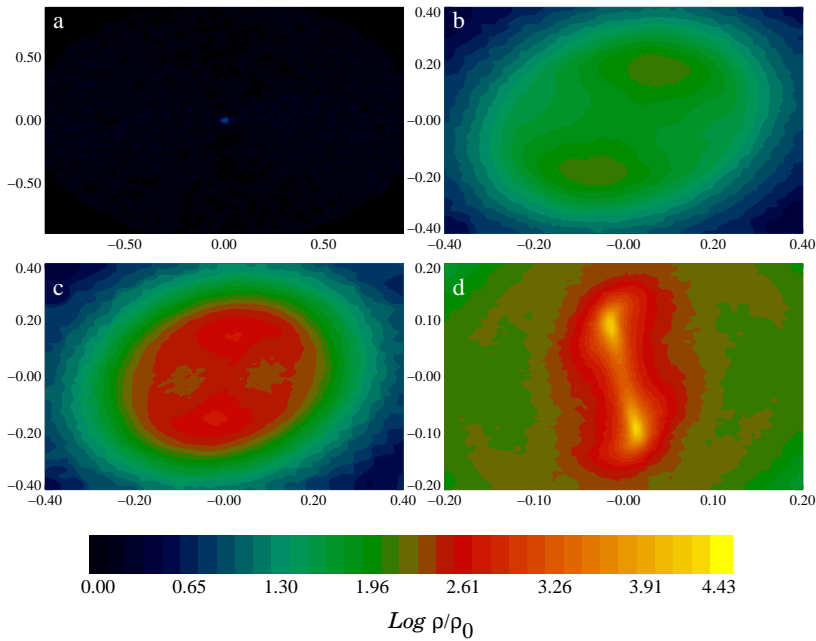}
\caption{\label{Mos4UnifRadr2_Ini} Iso-density plot to show the early
evolution of all the models.}
\end{center}
\end{figure}

\subsection{The radial mesh for core particles}
\label{subs:inicialcore}

We used an initial grid with spherical geometry, so that a
set of concentric shells were created and populated with
SPH particles in the following sense. We divided the total
volume $V$ of the sphere of radius $R_0$ into a
given number of bins, $N_{bin}$, such that $\Delta V=V/N_{bin}$ is
the volume of a spherical shell. Each shell can be characterized by
a radial interval $(r_l, r_f)$, such that its initial and final
radius are $r_l$ and $r_f$, respectively. The radius $r_f$ is
determined by the condition that $\Delta V$ be constant. Hence,
we have

\begin{equation}
r_f=\left( r_l + \frac{\Delta V}{4 \pi/3} \right)^{1/3}.
\label{finalradio}
\end{equation}
\noindent Thus, the first
shell is determined by the radial interval $(0,r_1)$, while the second shell
is delimited by $(r_1,r_2)$, and so on. Let us now define the average radius
$r_s$ and the radial width of a given shell as $r_s=(r_l+r_f)/2$ and 
$\delta_s= (r_f-r_l)$, respectively.

Then, by means of a Monte Carlo scheme, we populated each concentric shell with
a given number of equal mass particles, $N_{par}$, so that the particles were
located randomly in all the available surfaces of each spherical
shell. The spherical coordinates of the particles of a given shell
$(r_s,\theta,\phi)$ are related to uniform random variables $u$ and
$v$ (taking real values within the interval $[0,1]$) by the
following equations:

\begin{equation}
\label{eqcoord}
\begin{array}{l}
u = \frac{1-\cos(\theta)}{2} = \frac{1}{2}\int_0^{\theta}
\sin(\theta') \, d \theta' \, ,
\vspace{0.1 cm}\\
v = \frac{\phi}{2\pi} = \frac{1}{2 \, \pi }\int_0^{\phi} \, d \phi'
\end{array}
\end{equation}
\noindent We thus have a total of $N_{bin} \, \times \, N_{par}$ particles
distributed in the spherical volume of the core, such that the total
mass in each shell is constant and given by $N_{par} \, m_0$, where
$m_0$ is the particle mass, so that the global density of
the core is also constant. To achieve a constant density
distribution in a local sense, we further applied a radial
perturbation to all the particles of a given shell such that any particle
could be randomly displaced radially outward or inward, but preventing
a perturbed particle from reaching another shell. The radial 
perturbations $\epsilon_r$ applied to each SPH particle, regardless of the model, 
were at the order of $\epsilon_r = \delta_s / 2 $.

In the first panel of Fig.\ref{Mos4UnifRadr2_Ini}, one can appreciate
the spherical nature of the initial mesh, as only the innermost radial shell is
visible due to the huge contrast in density between this and the outer
shells.

In all the simulations of this paper, we used a total of two million SPH
particles, which according to the convergence study done by \cite{NuestroApJ},
is high enough to fulfill the resolution requirements described by
\cite{truelove}.

\subsection{Mass perturbation}
\label{subs:initialmasspert}

We ensured that a binary system would  be formed
in the simulation by implementing a mass perturbation
such that, if $m_0$ is the particle mass, the perturbed mass
$m_i$ of particle $i$ is $m_i=m_0+m_0*a \cos\left(m\, \phi_i \right)$,
where the perturbation amplitude is set to $a=0.1$ and the mode is
fixed at $m=2$; $\phi$ is the azimuthal spherical coordinate.

There are other methods for the implementation of density
perturbations, such as the Monte Carlo scheme, in which the particle mass
remains unchanged. Nevertheless, the mass perturbation we implemented here was
successfully applied in our previous papers on collapse; see
\cite{NuestroApJ} and \cite{NuestroRMAA}, and remarkably also for other
authors, such as \cite{springel}. With
particle mass variations within ten percent of the initial particle
mass, as is the case of our simulations, there are no border or particle
deficiency effects to worry about.

\subsection{The barotropic equation of state}
\label{subs:thermo}

To take into account the
heating of the gas due to both core contraction and energy
dissipation from artificial viscosity, we used the barotropic
equation of state proposed by~\cite{boss2000}:

\begin{equation}
p= c_0^2 \rho \left[ 1 +
\left( \frac{\rho}{\rho_{crit}}\right)^{\gamma -1 }
\right]
\label{beos}
\end{equation}
\noindent where $\gamma=5/3$ and $c_0$ is the sound speed so that
the corresponding temperature associated with the gas core is $T
\approx 10 \, $K. The critical density $\rho_{crit}$ determines the
change in the thermodynamic regime from isothermal to adiabatic. For
the early phases of the collapse, when
the peak density is much lower than the critical
density, $\rho_{max}<<\rho_{crit}$, the {\it beos} becomes an ideal equation of
state; for the late phases of the collapse,
when $\rho_{max}>>\rho_{crit}$, there is an increase in pressure
according to $p\approx \rho^{3/2}$, then the {\it beos} becomes an
adiabatic relation.

We use only one value given by $\rho_{crit}=5.0 \times 10^{-14} \,$ g
cm$^{-3}$. As we will show in the following sections, in the
simulations
considered in this paper, the average peak density reached
ranges around $\rho=2.5 \times 10^{-11} \,$ such that
the core density increased up to 3 orders of magnitude within the
adiabatic regime.

\section{Results}
\label{sec:resultados}
\begin{table}[ph]
\caption{
The models and their main results.}
{\begin{tabular}{|c|c|c|c|c|c|}
\hline
{Model} &  $M/M_{\odot}$  & $\beta$ & $t_{max}/t_{ff}$ & $ log_{10}
\left( \rho_{max}/\rho_0 \right) $
& Configuration \\
\hline
\hline
m075b0045 & 0.75 & 0.045 & 1.29 & 7.50 & Binary\\
\hline
m075b014  & 0.75 & 0.14  & 1.57 & 6.78  & Binary \\
\hline
\hline
m1b0045   & 1.0  & 0.045 & 1.11 & 7.28 & Binary \\
\hline
m1b014    & 1.0  & 0.14  & 1.38 & 6.79  & Binary \\
\hline
\hline
m1p5b0045 & 1.5  & 0.045 & 0.937& 6.99 & Primary \\
\hline
m1p5b011  & 1.5  & 0.11  & 1.08 & 6.94 & Binary \\
\hline
m1p5b014  & 1.5  & 0.14  & 1.10 & 6.70 & Binary \\
\hline
\hline
m2p5b0045 & 2.5 & 0.045 & 0.76 &  7.02 & Primary \\
\hline
m2p5b013  & 2.5 & 0.13 &  0.87  & 6.83 & Binary \\
\hline
m2p5b014 & 2.5 & 0.14  & 0.83  & 6.62 & Binary\\
\hline
\hline
m5b0045 & 5.0 & 0.045 & 0.53  & 6.58 & Primary \\
\hline
m5b014 & 5.0 & 0.14 & 0.71 & 7.73 & Primary \\
\hline
m5b021 & 5.0 & 0.21 & 0.71 & 6.47 & Binary \\
\hline
\end{tabular} 
\label{tab:modelos}}
\end{table}

\begin{figure}
\begin{center}
\begin{tabular}{c}
\includegraphics[width=2.25 in]{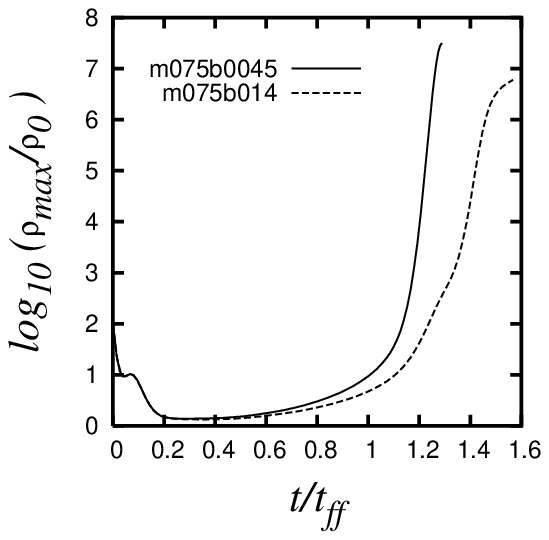}\\
\includegraphics[width=2.25 in]{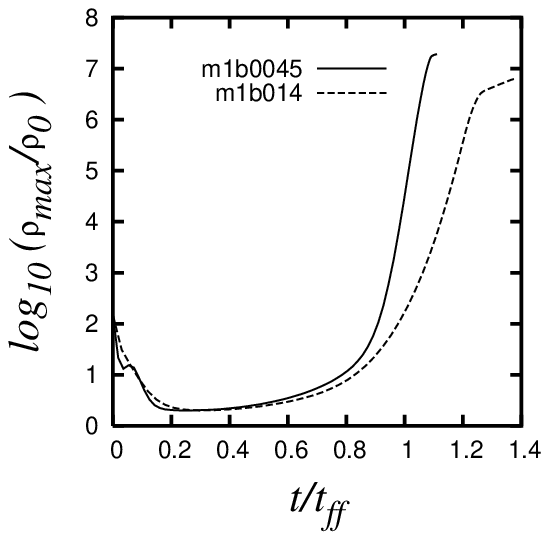} \\
\includegraphics[width=2.25 in]{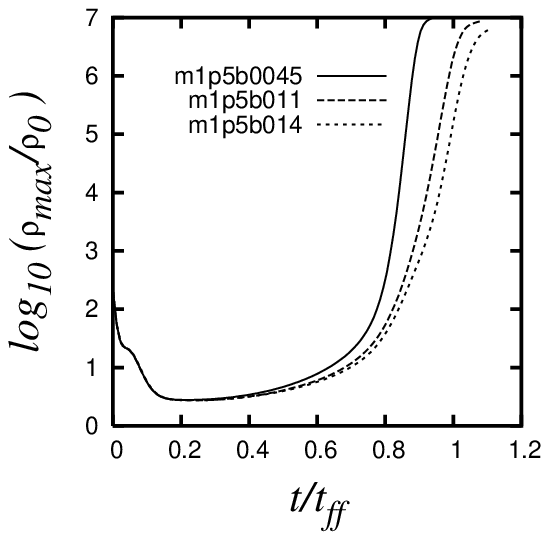}\\
\includegraphics[width=2.25 in]{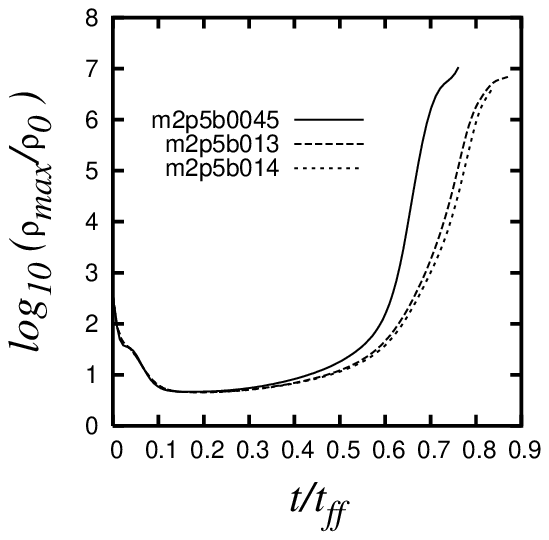}\\
\includegraphics[width=2.25 in]{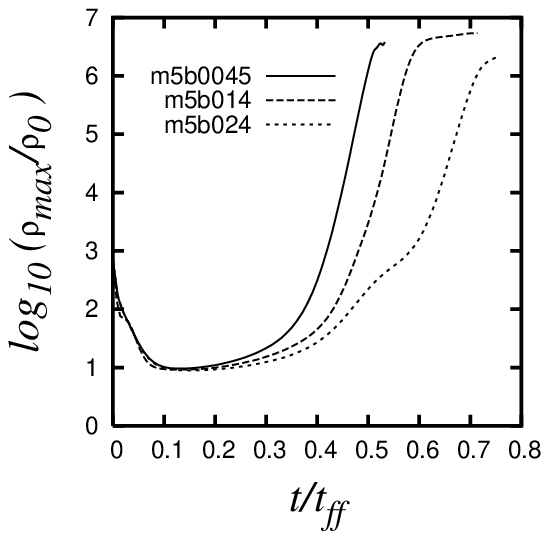}
\end{tabular}
\caption{ \label{DenMaxP} Time evolution of the peak density.}
\end{center}
\end{figure}

Worldwide simulations performed in order to follow the collapse of a uniform
density core have proven that an isolated rotating core contracts
itself to an almost flat configuration approximately within a
free-fall time of dynamical evolution; see for instance
\cite{bodeniv}, \cite{sigalotti2001} and the references therein.
Therefore, in order to illustrate our results, we used $2D$
iso-density plots for a slice of particles around the equatorial
plane of the core.

Let us now emphasize some important features of the early evolution
of the collapsing core, where the mass perturbation mentioned in
Section \ref{subs:initialmasspert} plays a fundamental role. When the peak density
reaches a value around $1.0 \,
\times 10^{-16}\,$g cm$^{-3}$, the mass perturbation generates
two well-defined mass condensations, which are clearly
visible in the second panel of Fig.\ref{Mos4UnifRadr2_Ini}. These
mass condensations act as mass attraction centers. As more mass is
being accreted by these centers, the peak density monotonically
increases on them and in their surroundings as well, as can be noted
by the color scale acquired by the central region of the core in
the third and fourth panels of Fig.\ref{Mos4UnifRadr2_Ini}, respectively. Thus,
all the models considered here finish this first evolution stage
with an embryonic binary system composed by two well-defined mass
condensations connected by a filament.

As can be seen in Fig.~\ref{DenMaxP}, our simulations easily satisfy
some basic expectations, some of which are: (i) all the models collapse by
the end of the simulation; (ii) the larger the value of $\beta$ given initially
to the core, the slower the core collapses; (iii) the more massive
the initial core, the faster it collapses; (iv) there must be a
maximum $\beta$ value, so that for $\beta > \beta_{max}$, 
the core simply expands without contracting.

First, there are two competing forces that
determine the next events to occur in the central core: on one hand,
the gravitational force, so that each mass condensation pulls on the
other, favoring their approach; on the other hand, centrifugal
force, acting on each mass condensation favoring their separation.

Now we shall separately illustrate the results of each model by
means of colored iso-density figures. In Table~\ref{tab:modelos} we
summarize the considered models and their main results, according to the
following entries: column one gives the label and column two
shows the total mass of the parent core; in column three we give the
$\beta$ initially provided to the core, while the maximum evolution
time and peak density of the simulation are shown in the fourth and
fifth columns, respectively; finally, the last column shows the
configuration obtained, either binary or primary, as explained below.
\begin{figure}
\begin{center}
\includegraphics[width=4.2in]{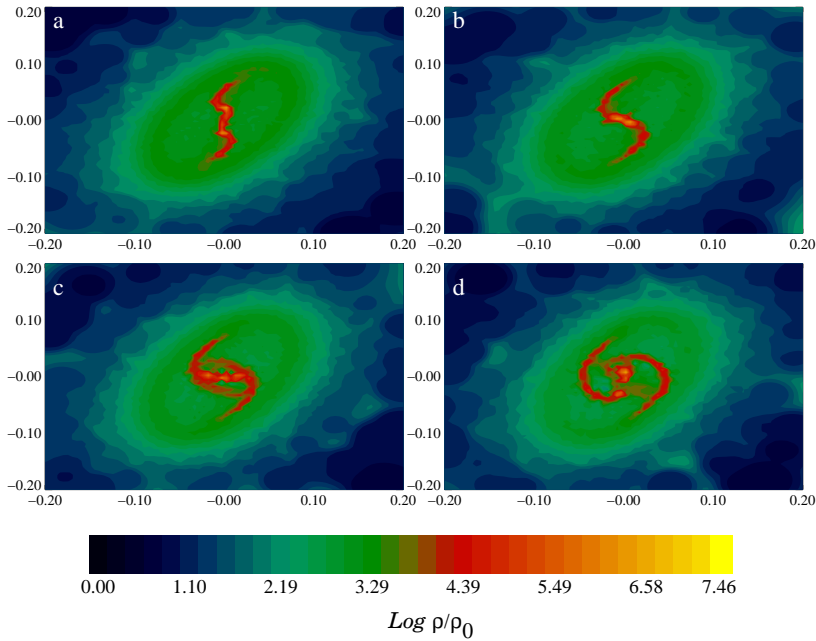}
\caption{\label{Mos4URadr6v1} Iso-density plot for model m0p75b0045.}
\end{center}
\end{figure}
\begin{figure}
\begin{center}
\includegraphics[width=4.2in]{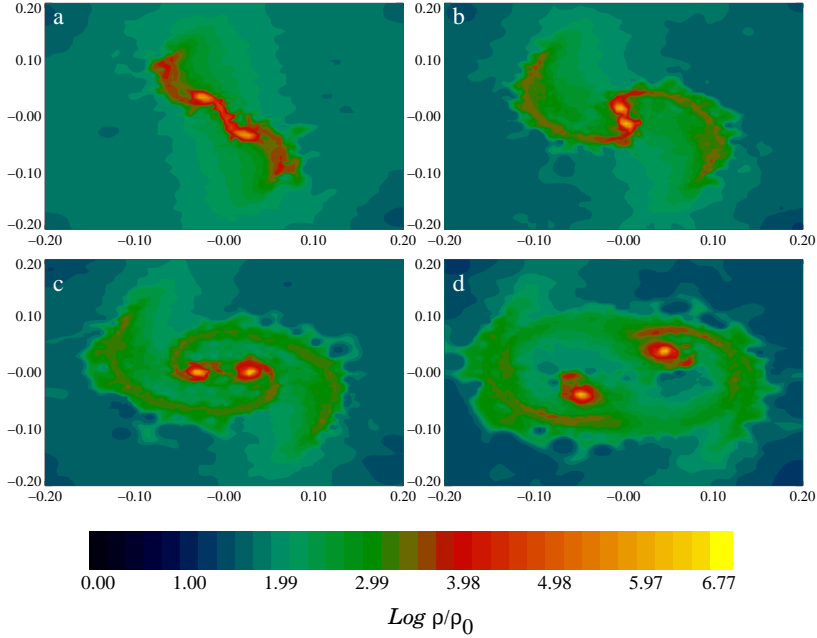}
\caption{\label{Mos4UnifRadr6} Iso-density plot for model m0p75b014.}
\end{center}
\end{figure}
\subsection{Models with parent core mass up to $M_0=1 \, M_{\odot}$}
\label{sub:m075ym1}

The models considered in this section are those labeled with m075 and
m1 in Table~\ref{tab:modelos}. For all these models, the
gravitational attraction between the formed mass condensations is
very easily overcome by the centrifugal repulsion; then the mass condensations
avoid the contact between them and fly apart to become truly fragments
which enter in orbit around one another, so we obtained the desired
binary configurations.

The iso-density plots for the low and high $\beta$ models with
$M_0=0.75 \, M_{\odot}$ can be seen in Fig.\ref{Mos4URadr6v1} and
Fig.\ref{Mos4UnifRadr6}, respectively. The binary separations are
around 111 AU and 402 AU, respectively. A clear mass asymmetry can
be seen between the fragments in the model m075b0045, as the masses
are $0.12 \, M_{\odot} $ and $0.03 \, M_{\odot}$, respectively.
Meanwhile, for model m075b014 the masses of the fragments are almost
the same, around $0.07 \, M_{\odot}$.

\begin{figure}
\begin{center}
\includegraphics[width=4.2in]{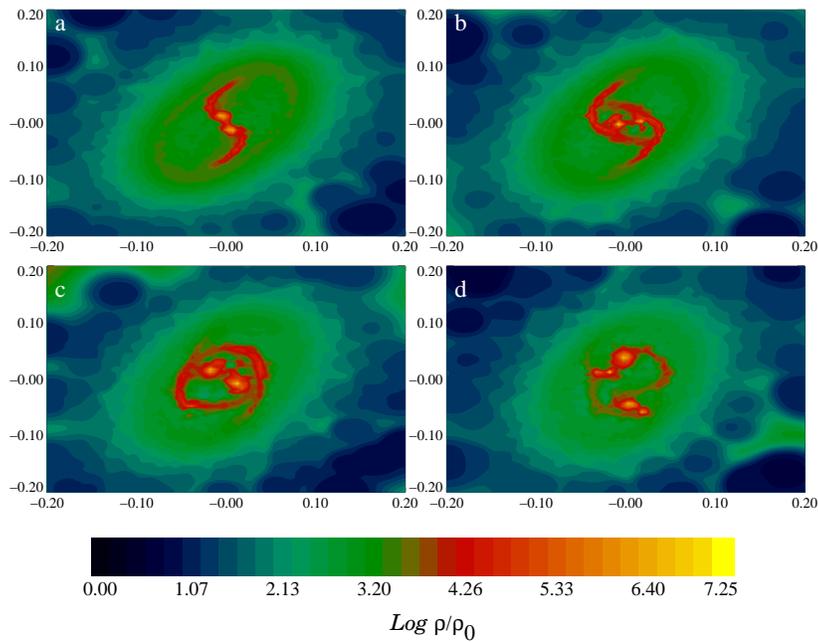}
\caption{\label{Mos4URadr2v1} Iso-density plot for model m1b0045.}
\end{center}
\end{figure}

\begin{figure}
\begin{center}
\includegraphics[width=4.2in]{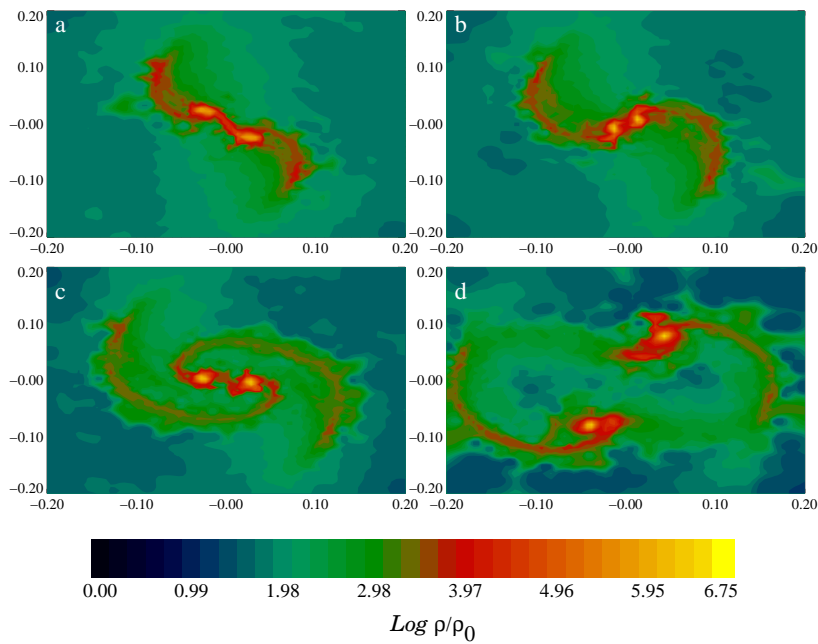}
\caption{\label{Mos4UnifRadr2} Iso-density plot for model m1b014.}
\end{center}
\end{figure}

Now, let us consider the results for the low and high $\beta$
models with $M=1 \, M_{\odot}$, which are shown in
Fig.\ref{Mos4URadr2v1} and Fig.\ref{Mos4UnifRadr2}, respectively.
The corresponding binary separations have now increased up to 326 and 667
AU, respectively. The masses of the fragments for model m1b0045 are
$0.16 \, M_{\odot} $ and $0.09 \, M_{\odot}$, while for model m1b014 the
masses are very similar between them, around $0.11 \, M_{\odot} $.

Thus, we see that a small change in the total mass of the
parent core produces a very large change in the resulting binary separation
and mass. As expected, in the low $\beta$ models, the separation reached by the
mass condensations is smaller than for the high $\beta$ models.

\subsection{Models with parent core mass $M_0=1.5 \, M_{\odot}$}
\label{sub:m1p5}

\begin{figure}
\begin{center}
\includegraphics[width=4.2in]{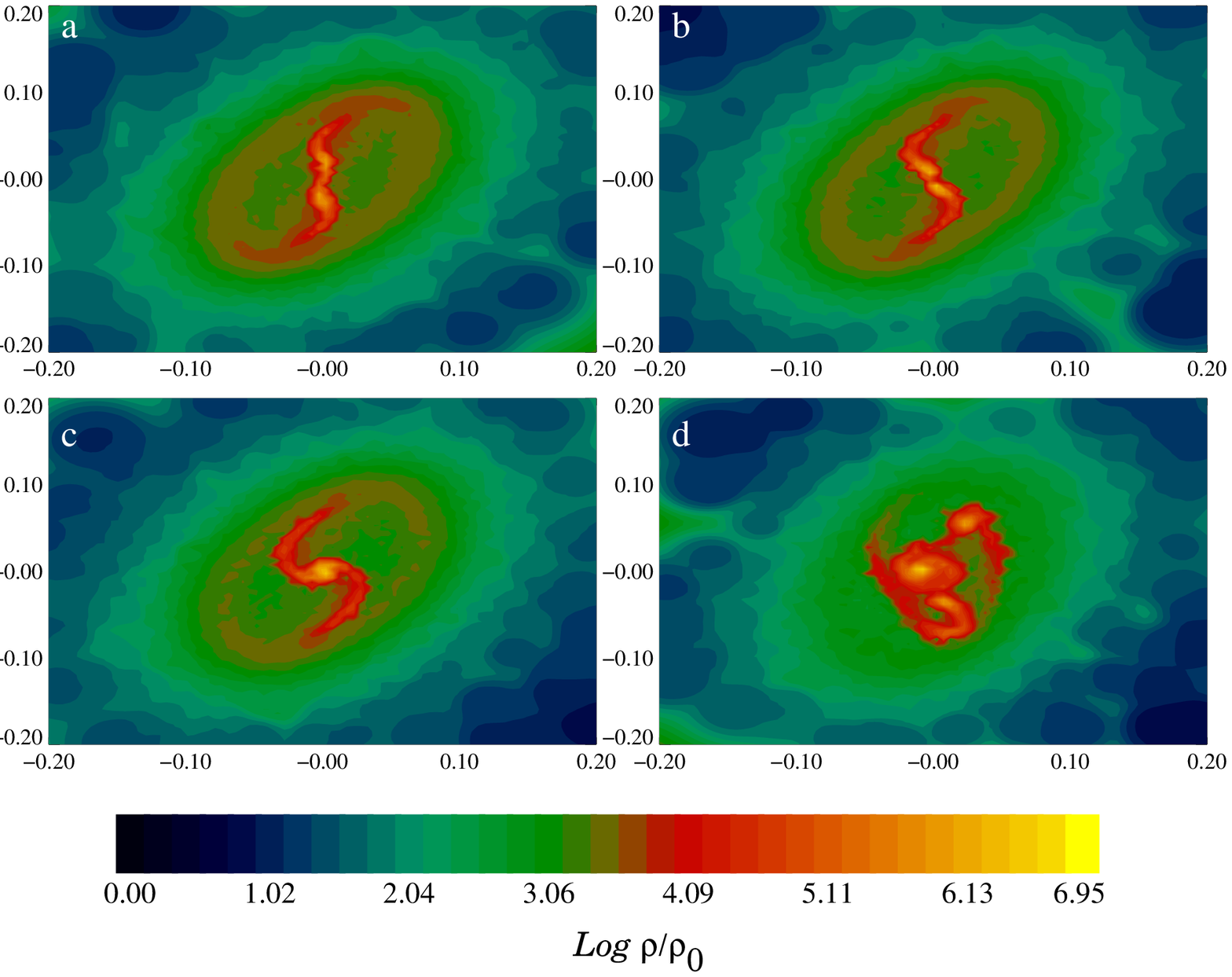}
\caption{\label{Mos4URadr5v1} Iso-density plot for model m1p5b0045.}
\end{center}
\end{figure}

For the first time we saw that the low $\beta$ model m1p5b0045 did
not produce a binary via the separation of its
embryonic mass condensations, and instead we saw their merging. So,
only a primary mass condensation is formed in the central core
region, which is surrounded by small spiral arms, as can be seen in
Fig.\ref{Mos4URadr5v1}. Soon thereafter, these spiral arms break and
separate from the primary, so the simulation ended with a
primary mass accompanied by two smaller mass condensations.

\begin{figure}
\begin{center}
\includegraphics[width=4.2in]{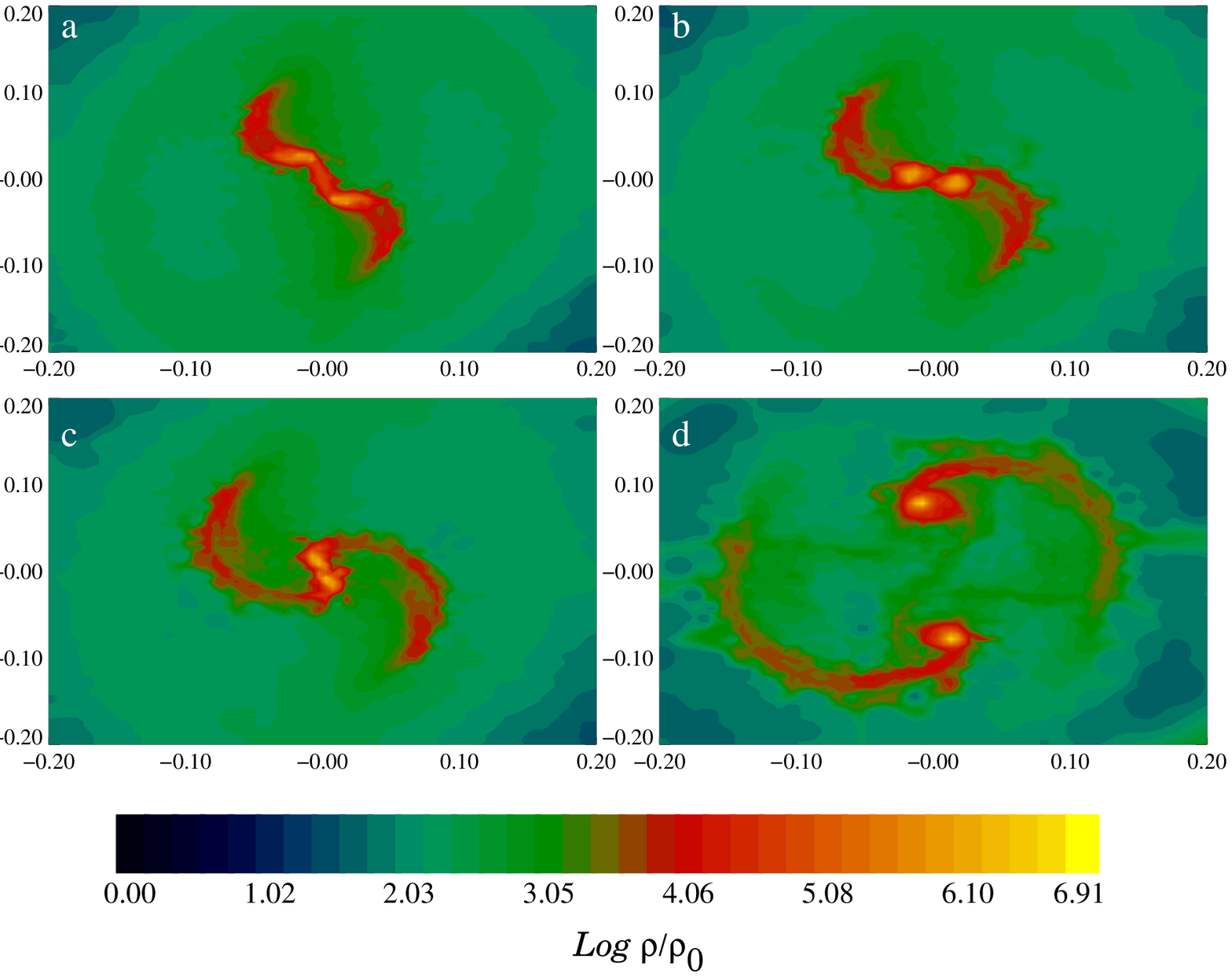}
\caption{\label{Mos4URadr5v1p3} Iso-density plot for model m1p5b011.}
\end{center}
\end{figure}

According to our strategy, we then increased the
angular velocity of this model up to the value where we got $\beta=0.11$, so
that we had now the model m1p5b011, in which we again obtained the appearance
of a binary system via the separation of the embryonic mass condensations; see
Fig.\ref{Mos4URadr5v1p3}. The binary separation in this case, 527 AU, is
similar to that already seen in Section\ref{sub:m075ym1} for the low $\beta$
model m1b014. The masses of the fragments for this new model
are $0.16\, M_{\odot} $ and $0.17 \, M_{\odot}$.

\begin{figure}
\begin{center}
\includegraphics[width=4.2in]{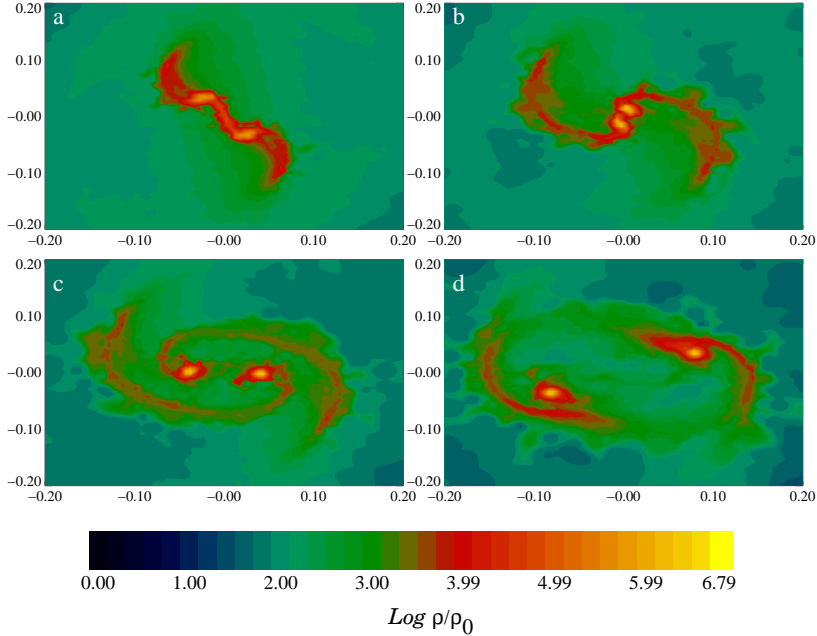}
\caption{\label{Mos4UnifRadr7} Iso-density plot for model m1p5b014.}
\end{center}
\end{figure}

The existence of model m1p5b011 tells us in advance that
the high $\beta$ model will form the desired binary, as can be seen
in Fig.\ref{Mos4UnifRadr7}, where we show the results for model
m1p5b014. As was previously observed, the additional rotational energy
produced a small increase in the binary separation, as we now obtained
585 AU, while the masses of the fragments were almost identical
at $0.16\, M_{\odot} $.


\subsection{Models with parent core mass $M_0=2.5 \, M_{\odot}$}
\label{sub:m2p5}

\begin{figure}
\begin{center}
\includegraphics[width=4.2in]{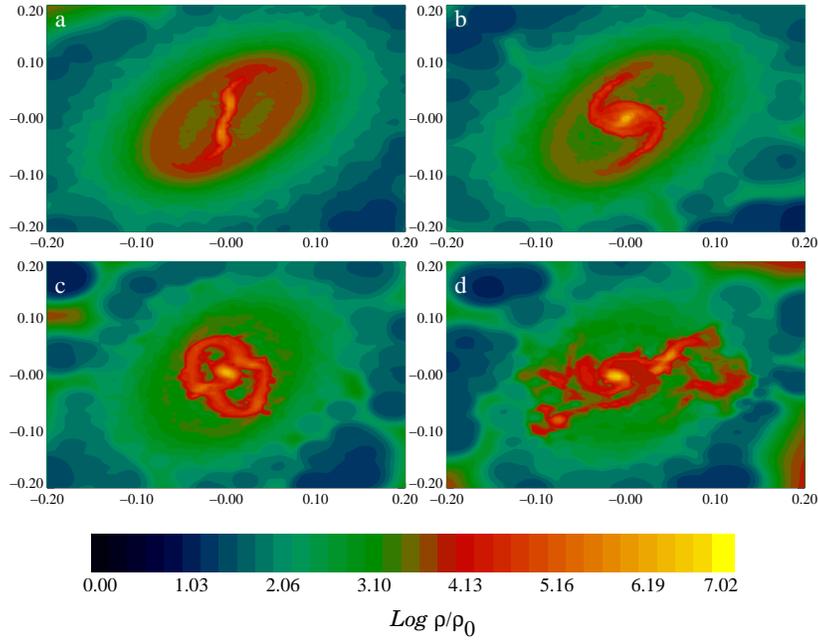}
\caption{\label{Mos4URadr3v1} Iso-density plot for model m2p5b045.}
\end{center}
\end{figure}

\begin{figure}
\begin{center}
\includegraphics[width=4.2in]{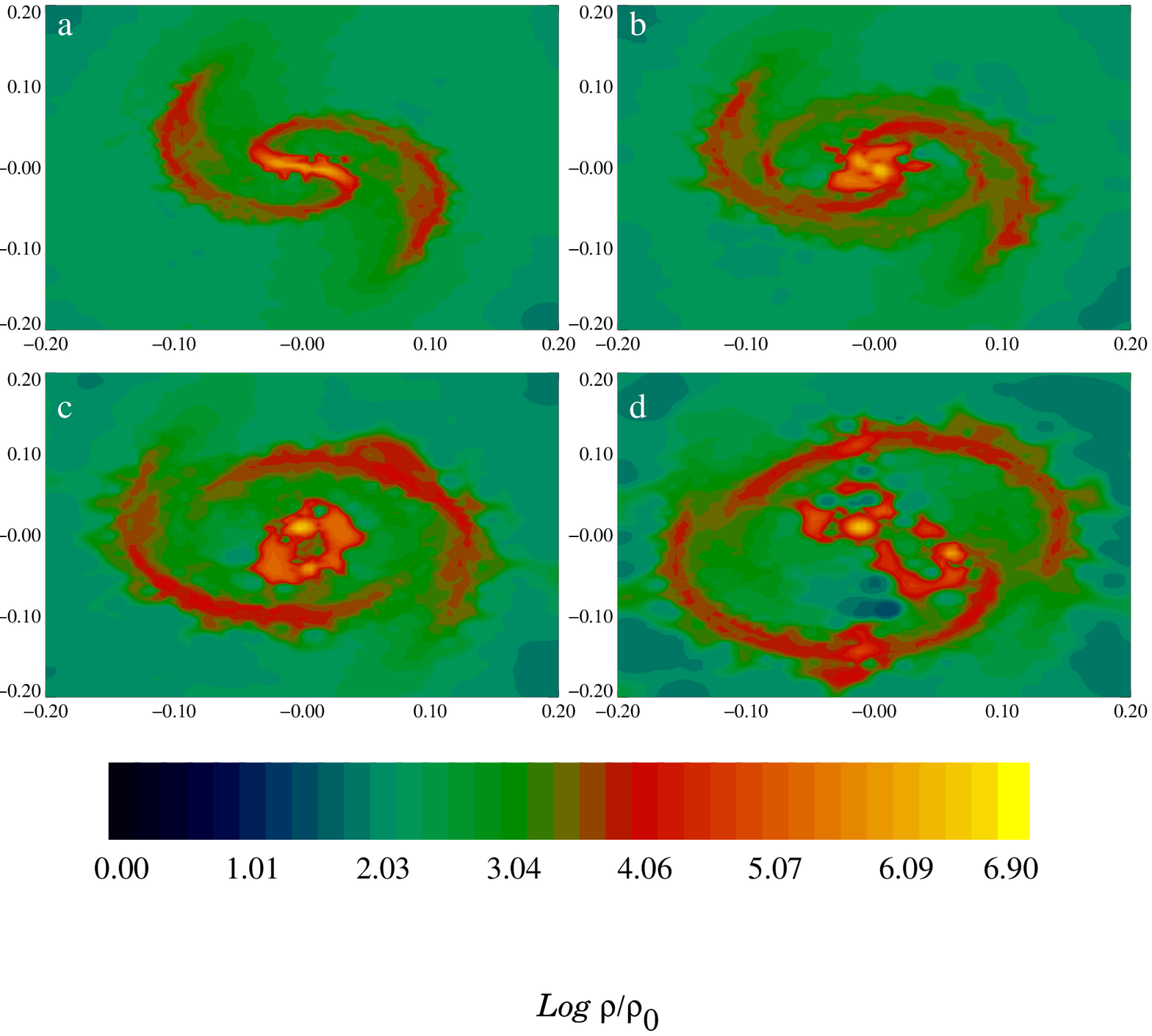}
\caption{Iso-density plot for model m2p5b013.}
\label{Mos4URadr3v1p5}
\end{center}
\end{figure}

\begin{figure}
\begin{center}
\includegraphics[width=4.2in]{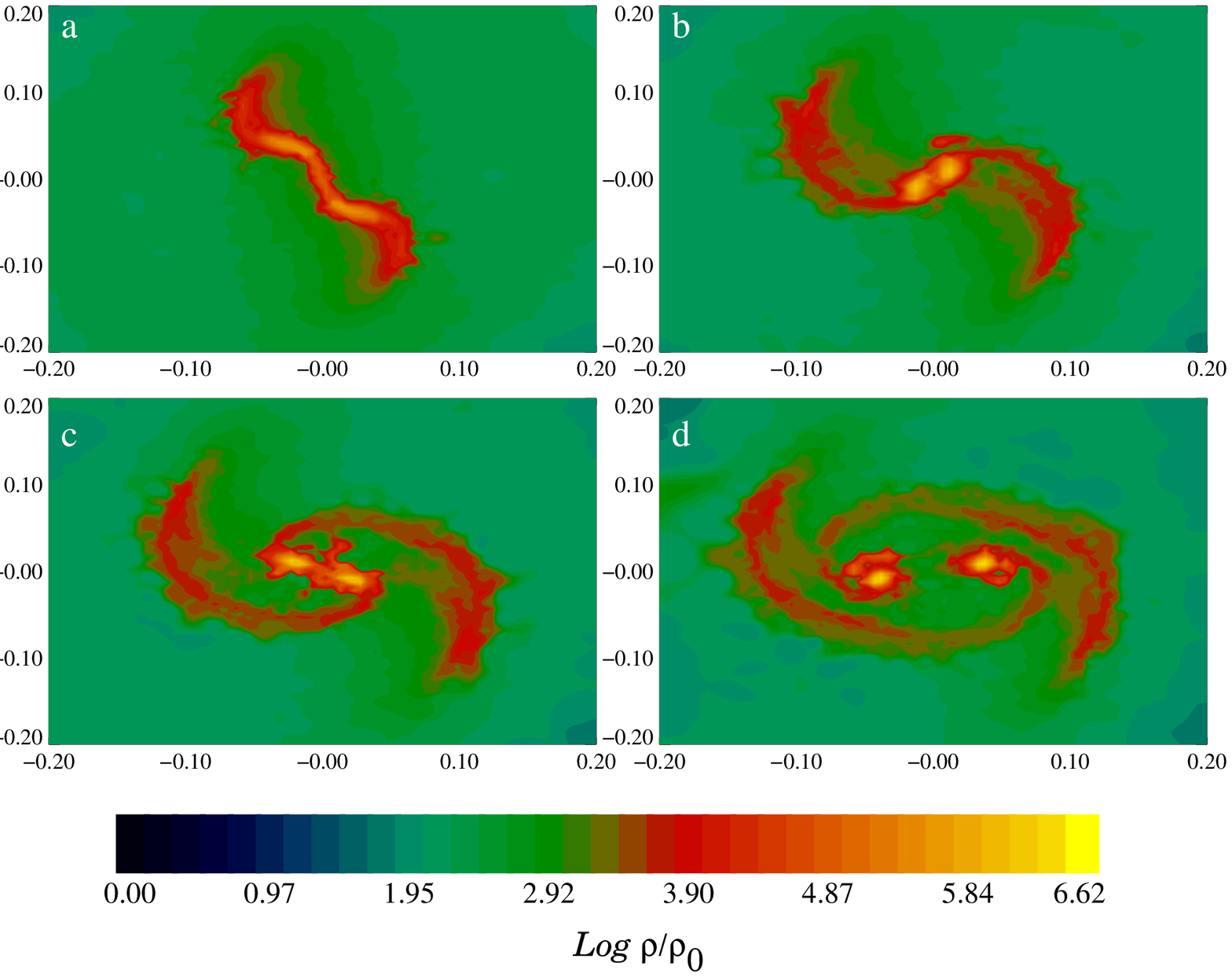}
\caption{\label{Mos4UnifRadr9} Iso-density plot for model m2p5b014.}
\end{center}
\end{figure}

As was the case in Section~\ref{sub:m1p5}, here the low $\beta$ model
m2p5b0045 produced a primary configuration, which is shown in
Fig.\ref{Mos4URadr3v1}. However, the high $\beta$ model m2p5b014 produce
the desired binary configuration. For this reason, we
expected to find a new $\beta$ value, such that the model m2p5b0045
would become a binary system. So, by systematically increasing
its $\beta$, we reached the value $\beta=0.13$, where we found the
desired configuration; see Fig.\ref{Mos4URadr3v1p5}. The results for
model m2p5b014 are shown in Fig.\ref{Mos4UnifRadr9}.

One would expect to find very similar physical properties for
these pair of models, m2p5b013 and m2p5b014, as their initial $\beta$
values are very similar. The binary separations are indeed very similar,
around 259 and 262 AU, respectively. However, the masses of their
fragments are not so similar. This mass difference can be explained
by closely looking Fig.\ref{Mos4URadr3v1p5}, where one can
notice that there is an important mass exchange between the
mass condensations, as the additional rotational energy supplied is
perhaps barely enough to separate them, so they do not become true fragments.

\subsection{Models with parent core mass $M_0=5 \, M_{\odot}$}
\label{sub:m5}

\begin{figure}
\begin{center}
\includegraphics[width=4.2in]{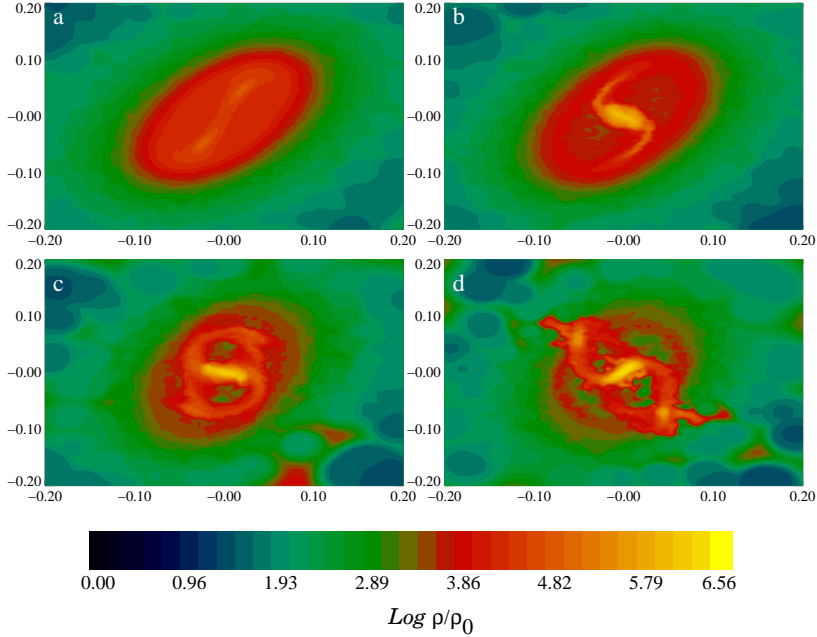}
\caption{ \label{Mos4URadr7v1} Iso-density plot for modelm5b045.}
\end{center}
\end{figure}

\begin{figure}
\begin{center}
\includegraphics[width=4.2in]{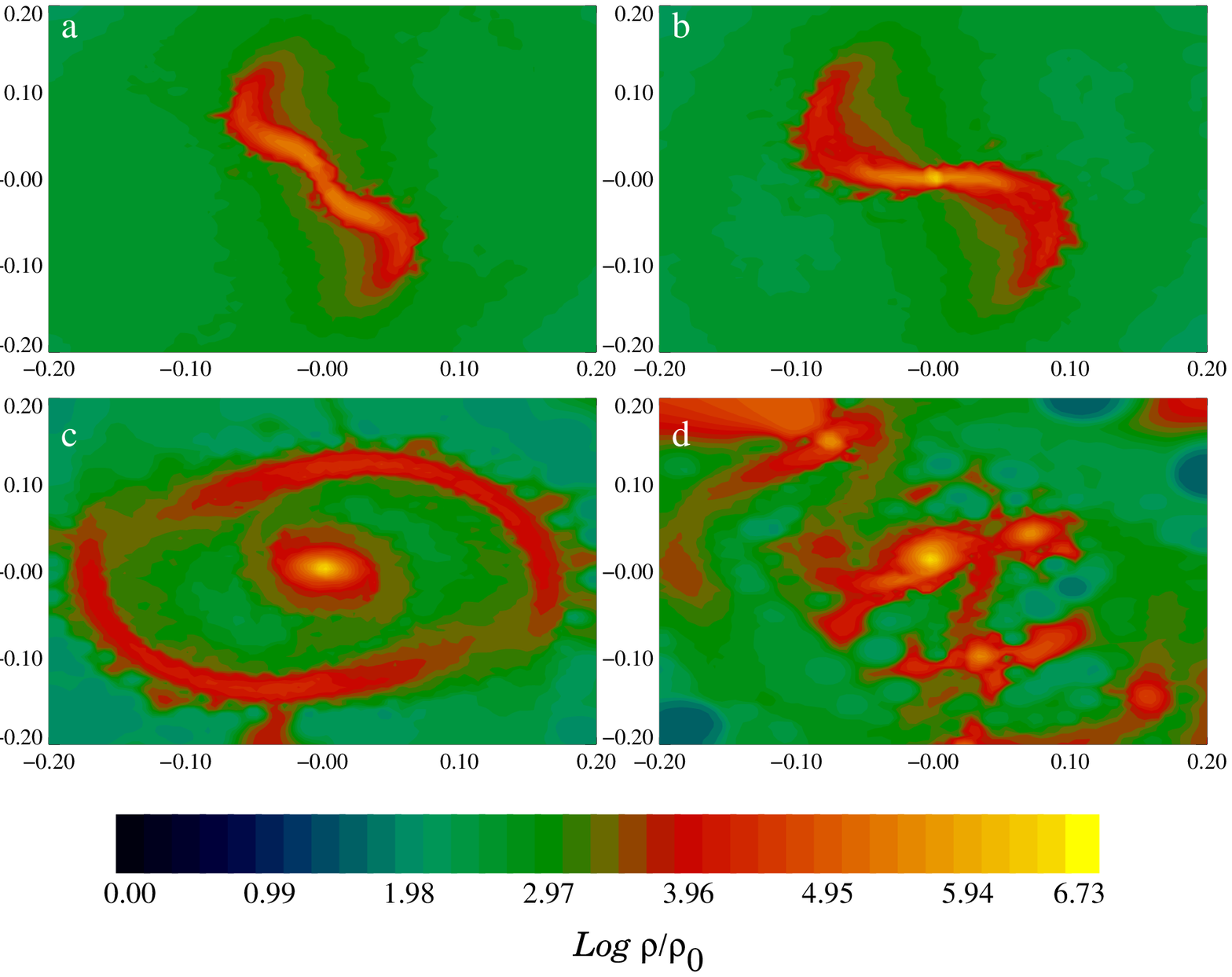}
\caption{ \label{Mos4UnifRadr5} Iso-density plot for model m5b014.}
\end{center}
\end{figure}

In this case we observed for the first time that even the high
$\beta$ model does not produce the desired binary configuration. But
we can still compare models m5b0045 and m5b014, as they form two
slightly different primary dominated configurations. In model
m5b0045, an elongated central bar is formed with two additional mass
condensations formed at the ends of the spiral arms, as
illustrated in Fig.\ref{Mos4URadr7v1}. In model m5b014 we see
again the formation of a central mass condensation, but in this case,
it is surrounded by very long spiral arms, which break soon thereafter and
separate from this central mass; see Fig.\ref{Mos4UnifRadr5}.

\begin{figure}
\begin{center}
\includegraphics[width=4.2in]{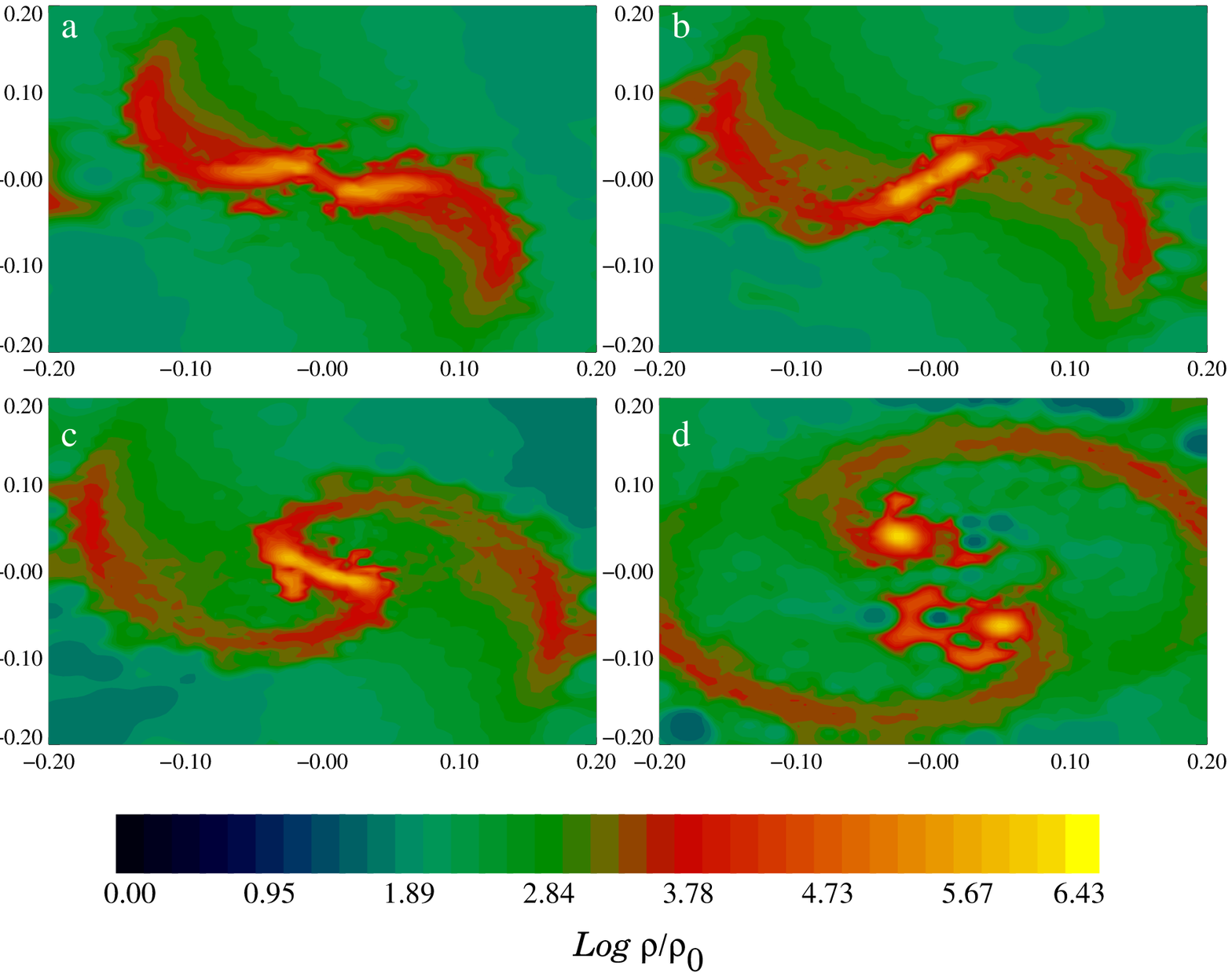}
\caption{ \label{Mos4UnifRadr5p4} Iso-density plot for model m5b021.}
\end{center}
\end{figure}

As usual, we then increased the level of the initial rotational energy, until
the value of $\beta=0.21$, where we obtained the desired binary
configuration, labeled now as model m5b021; see Fig.\ref{Mos4UnifRadr5p4}. The
binary separation is 427 AU while the masses of the fragments are
$0.6 \, M_{\odot} $ and $0.4 \, M_{\odot}$.
\section{Discussion}
\label{sec:dis}

\begin{table}[ph]
\caption{Physical properties of fragments.}
{\begin{tabular}{|c|c|c|c|c|}
\hline
{Model} &  $r_{max}/R_0$ & $M_f/M_{\odot}$  & $\alpha_f$ & $\beta_f$  \\
\hline
\hline
m075b0045 & 0.0125 &  1.2286294e-01 & 0.266161  &  0.181729 \\
\hline
m075b0045 & 0.0125 &  3.2744151e-02 & 0.207690  &  0.261402 \\
\hline
\hline
m075b014  & 0.0125 &  7.3241010e-02 & 0.202020  &  0.244268 \\
\hline
m075b014  & 0.0125 &  7.0384055e-02 & 0.204878  & 0.237756 \\
\hline
\hline
m1b0045   & 0.0164 &  1.6832440e-01 &  0.233851 &  0.265091 \\
\hline
m1b0045   & 0.0164 &  9.3557134e-02 &  0.276662 &  0.213533 \\
\hline
\hline
m1b014    & 0.0188 &  1.1661998e-01 &  0.207951 &  0.225678\\
\hline
m1b014    & 0.0188 &  1.1168584e-01 &  0.207141 &  0.243163\\
\hline
\hline
m1p5b011  & 0.028  & 1.6871537e-01  &  0.255228 &  0.171645 \\
\hline
m1p5b011  & 0.028  & 1.7556763e-01  &  0.244072 &  0.195569\\
\hline
\hline
m1p5b014  & 0.028  & 1.6674188e-01  &  0.234014 &  0.243616 \\
\hline
m1p5b014  & 0.028  & 1.6345865e-01  &  0.221192 &  0.234322\\
\hline
\hline
m2p5b013 & 0.0188  & 3.7504950e-01  & 0.239499 &  0.196338\\
\hline
m2p5b013 & 0.0188  & 1.0161296e-01  & 0.256503 & 0.124867 \\
\hline
m2p5b014 & 0.0188  & 2.7242526e-01  &  0.246773 &  0.222628\\
\hline
m2p5b014 & 0.0188  & 2.6820856e-01  &  0.242448 &  0.205314\\
\hline
\hline
m5b021   & 0.0329  & 6.3345021e-01  &  0.249588 &  0.174287\\
\hline
m5b021   & 0.0329  & 4.0061280e-01  &  0.262765 &  0.159544 \\
\hline
\hline
\end{tabular} 
\label{tab:properties}} 
\end{table}
\begin{table}[ph]
\caption{
Binary separation and the sound speed.}
{\begin{tabular}{|c|c|c|} \hline
{Model} &  $r_{sep} [AU]$ & $c_0$ [cm/s] \\
\hline
\hline
m075b0045 & 111.43  &  13760.78 \\
\hline
m075b014  & 402.98  &  13799.62 \\
\hline
\hline
m1b0045   & 326.7 & 16647.83 \\
\hline
m1b014    &  666.79 & 16647.83 \\
\hline
\hline
m1p5b011  & 527.92 & 20000.0 \\
\hline
m1p5b014  & 585.93 & 19711.58 \\
\hline
\hline
m2p5b013 & 259    &  25410.62 \\
\hline
m2p5b014 & 262.2 &  25410.62 \\
\hline
\hline
m5b021 & 427.97 &  35820.06 \\
\hline
\end{tabular} 
\label{tab:sep} }
\end{table}

\begin{figure}
\begin{center}
\includegraphics[width=4.2in]{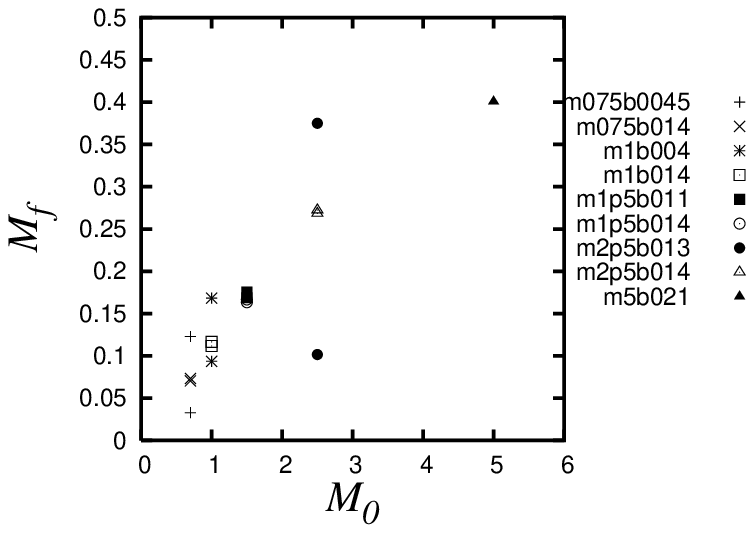}
\caption{\label{MasaFragvsM} Binary mass given in terms of $M_{\odot}$.}
\end{center}
\end{figure}

The main results of this paper are already contained in
Tables~\ref{tab:modelos}, \ref{tab:properties} and \ref{tab:sep}.
However, it is more illustrative to present them visually, so
in Figs.\ref{Radr_Ejes}, \ref{MasaFragvsM}, \ref{CentrosFrag_UltC_Todos_AU} and
\ref{Vel3D}, we show: (i) the
schematic diagram where the desired binary configurations are located; (ii)
the obtained binary mass $M_f$ as a function of the total mass of
the core $M_0$; (iii) the obtained binary separations and (iv) the distribution
of the velocity field. Let us now comment upon the creation of each figure
and their main results.

\subsection{The schematic diagram of binary configurations}
\label{subsec:diag}

\begin{figure}
\begin{center}
\includegraphics[width=4.2in]{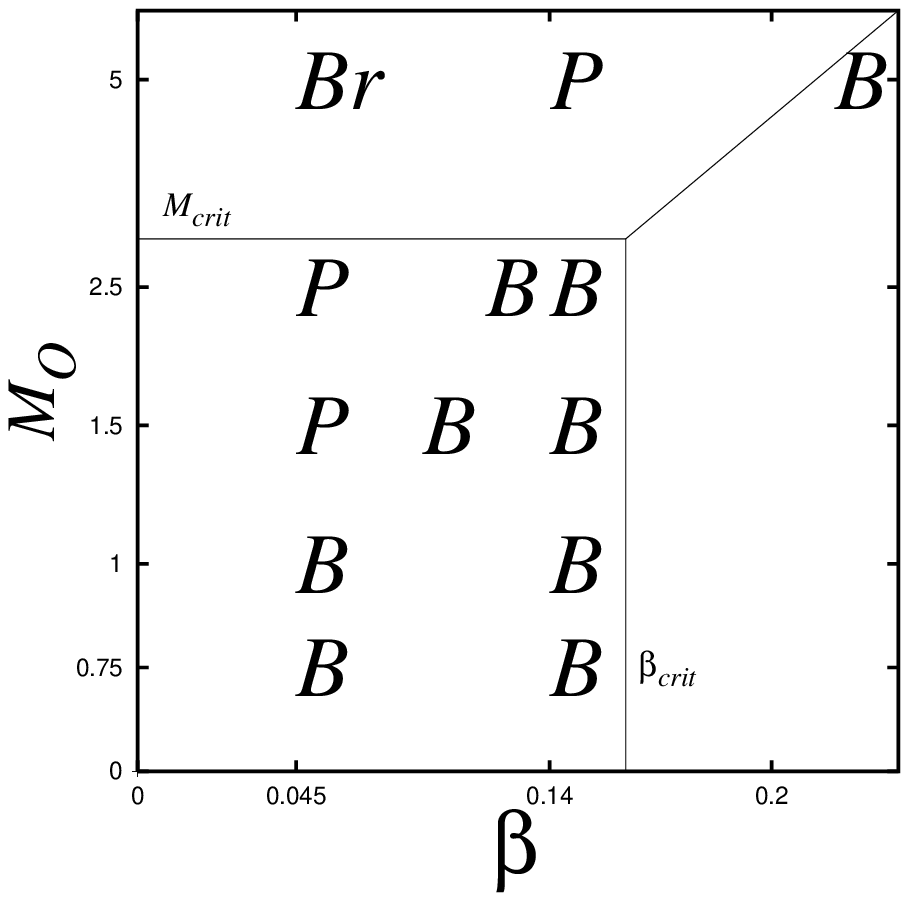}
\caption{ \label{Radr_Ejes}Schematic diagram to show the location of the
binary systems. The mass $M_0$ is given in terms of $M_{\odot}$.}
\end{center}
\end{figure}

The calculated schematic diagram of desired binary configurations is
illustrated in Fig.\ref{Radr_Ejes}. We first notice that there must
exist a critical mass $M_{crit}$ that corresponds to a
$\beta_{crit}$, such that they separate two regimes: one where $M_0
> M_{crit}$, in which the needed $\beta$ to obtain the desired
binary configuration must be higher than the $\beta_{crit}$, but
smaller than the maximum $\beta_{max}$ that allows to the core to
remain in a bounded configuration, that is, $\beta$ is within the
interval $\beta \, \in \, (\beta_{crit}, \beta_{max})$.

Another regime, with $M_0 < M_{crit}$, in which the needed $\beta$ can take
values in the interval $ 0 < \beta <\beta_{crit}$, where one can
definitely obtain the desired binary configuration. However, there is
still the possibility of having a binary
configuration with even higher values of
$\beta \, \in \,(\beta_{crit}, \beta_{max})$, as
the high value $\beta=0.14$ is set arbitrarily.

The cores were observed to have low
rotational velocities, so if we had a collapsing core with
$M_0<M_{crit}$, then according to Fig.\ref{Radr_Ejes}, it would be
more likely to have a binary configuration resulting from its
collapse. On the contrary, if we had a collapsing core with
$M_0>M_{crit}$, then it would be more likely to have a primary
system as the result of its collapse.

\subsection{The physical properties of binaries}
\label{subsec:corre}

Let us now consider Fig.\ref{MasaFragvsM}, recalling first that, as
an approximation strategy to construct our simulation models, in this
paper we increased the core mass $M_0$ without changing the
core radius $R_0$, so the core density is increased
and therefore the free fall time is decreased; see Eq.\ref{tff}.

The accreting mass rate can
be inaccurately estimated by means of $\dot{M}= M_0/t_{ff}$; that is, as if
the entire core mass had collapsed in
a free fall time. The combinations of changes mentioned above, that the $M_0$
increases while the $t_{ff}$ decreases, gives us an
increasing $\dot{M}$ for all the models under consideration.

A better estimate for the accreting mass rate was obtained by a
semi-analytical approach to the collapse of an isothermal core; see
\cite{shu}, which is given now by $\dot{M}=c_0^3/G$,  where $c_0$ is the
sound speed and $G$ is Newton's gravitational constant. In order to
keep the ratio of thermal energy to gravitational energy,
denoted by $\alpha$, fixed in all our models, we increased the sound
speed in our models, so that one sees that the $\dot{M}$ increases
when the mass of the core increases at least for the first stage of
evolution where the isothermal approximation is valid.

So, the increase of the mass of the binary fragment with increase of the mass
of the parent core is expected, as the mass accretion rate also
increases with the mass of the parent core. In this paper we confirmed
this expectation and we measured the binary mass $M_f$ obtained
out of a given initial $M_0$. It should be noted that this fragment
mass $M_f$ was determined in the following way: first we took the
highest density particle in the region where the fragment was
located. This particle is considered the center of the fragment. We
then found all the SPH particles whose density was greater than or
equal to some minimum density value given in advance by $
\log_{10} \left( \rho_{min}/\rho_0 \right)=5.0$ and that were within
a given maximum radius $r_{max}$ from the fragment center; see the
second column of Table~\ref{tab:properties}. These parameters
correspond to a minimum density of $3.82 \, \times 10^{-13}\,$g cm$^{-3}$
and a maximum radius in the range of $r_{max}=41-100$ AU.

This set of particles defined the fragment and allowed us to
calculate its integral properties; for instance, its mass $M_f$, and
their ratios $\alpha_f$ and $\beta_f$. These calculated integral
properties are shown in columns 3, 4 and 5 of
Table~\ref{tab:properties}, respectively, in which the number of
selected particles falls in the range of 150 to 300 thousand,
approximately.

In Section~\ref{sec:resultados}, we mentioned the calculated mass
of the binary configurations. It is still necessary to comment on
the calculated energy ratios $\alpha_f$ and $\beta_f$. It was
demonstrated by \cite{SegRMAA} that in general the fragments
obtained out by the fragmentation of a rotating core tend to
virialize. We observed that for the last snapshot available in
each simulation the sum of $\alpha_f$ and $\beta_f$ was always
less than 0.5, so the fragments were still collapsing.

\begin{figure}
\begin{center}
\includegraphics[width=4.2in]{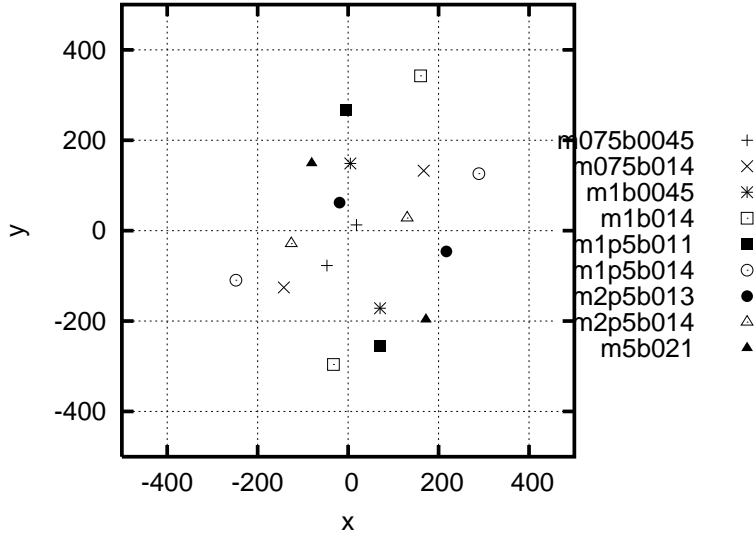}
\caption{\label{CentrosFrag_UltC_Todos_AU} Binary separation given in AU.}
\end{center}
\end{figure}

Finally, in order to determine the binary separations illustrated in
Fig.\ref{CentrosFrag_UltC_Todos_AU}, we simply calculate the
distance between the centers associated with each fragment, as
defined according to the procedure outlined in the previous
paragraph. It is important to mention that the longest binary separations
were obtained for intermediate mass models: the m1b014 and m1p5b011;
see Table~\ref{tab:sep}.

\subsection{The velocity distribution of binaries}
\label{subsec:distri}

\begin{figure}
\begin{center}
\begin{tabular}{cc}
 & \includegraphics[width=3.2in]{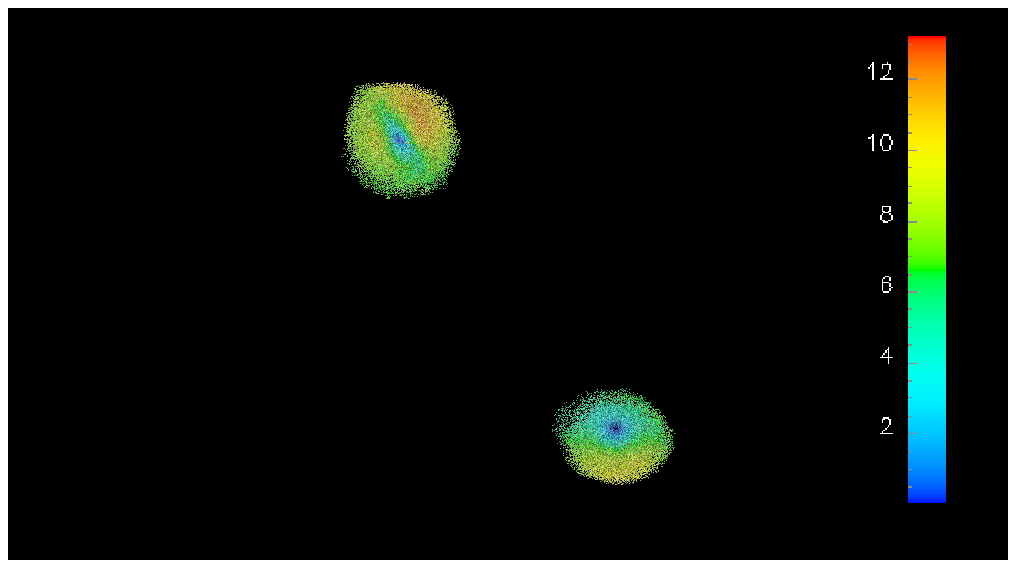} \\
\includegraphics[width=3.2in]{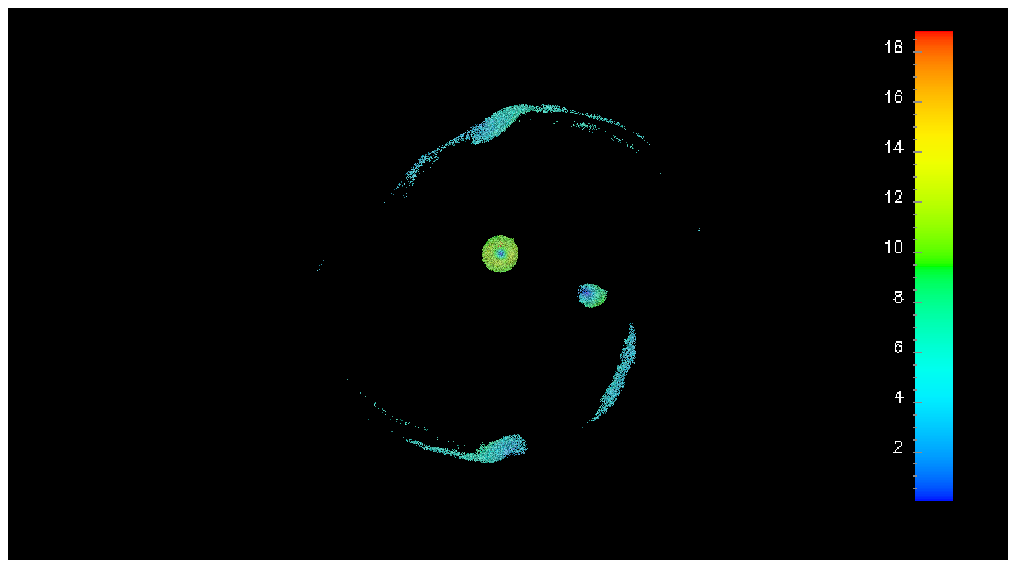} &
\includegraphics[width=3.2in]{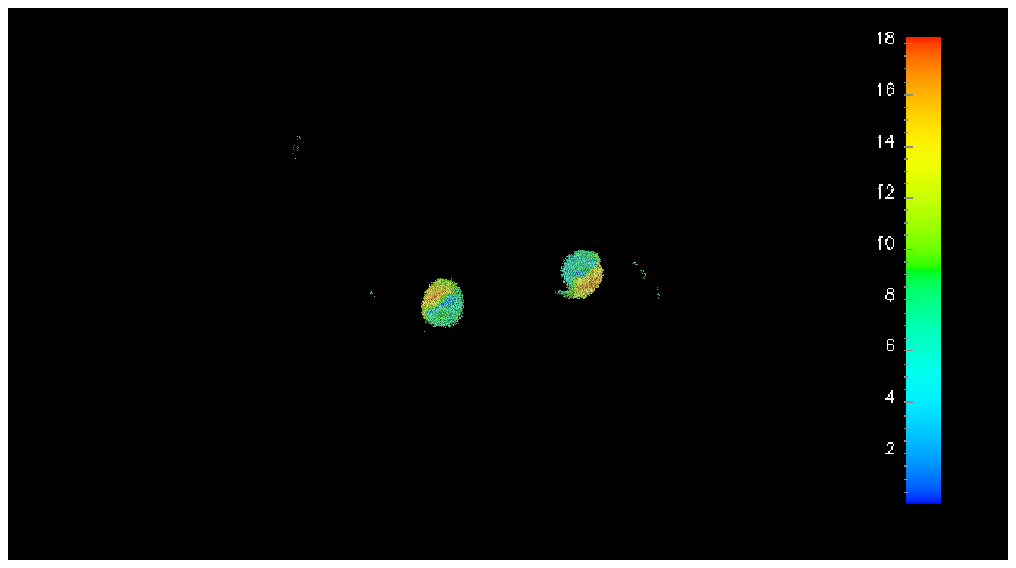} \\
\includegraphics[width=3.2in]{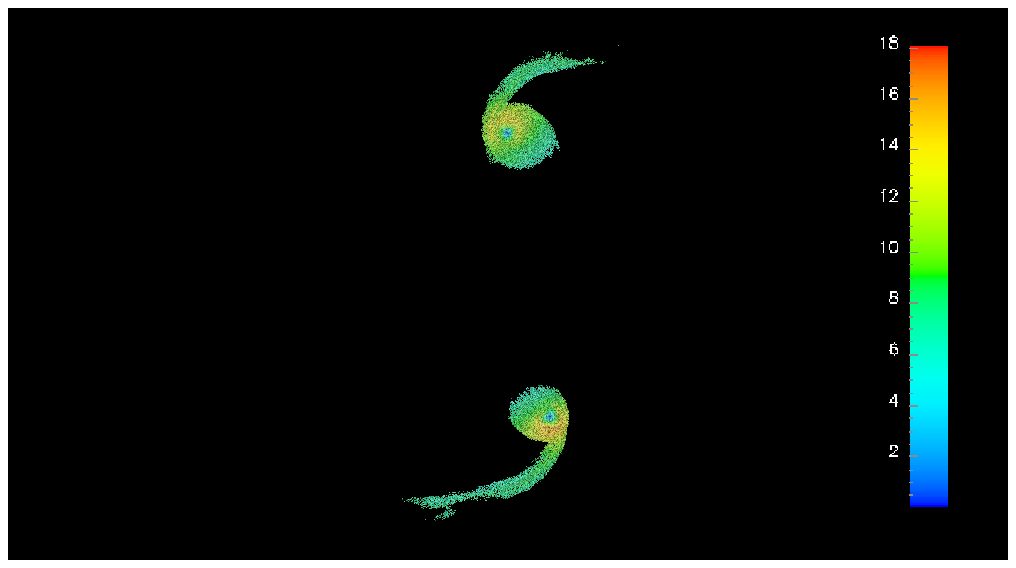} &
\includegraphics[width=3.2in]{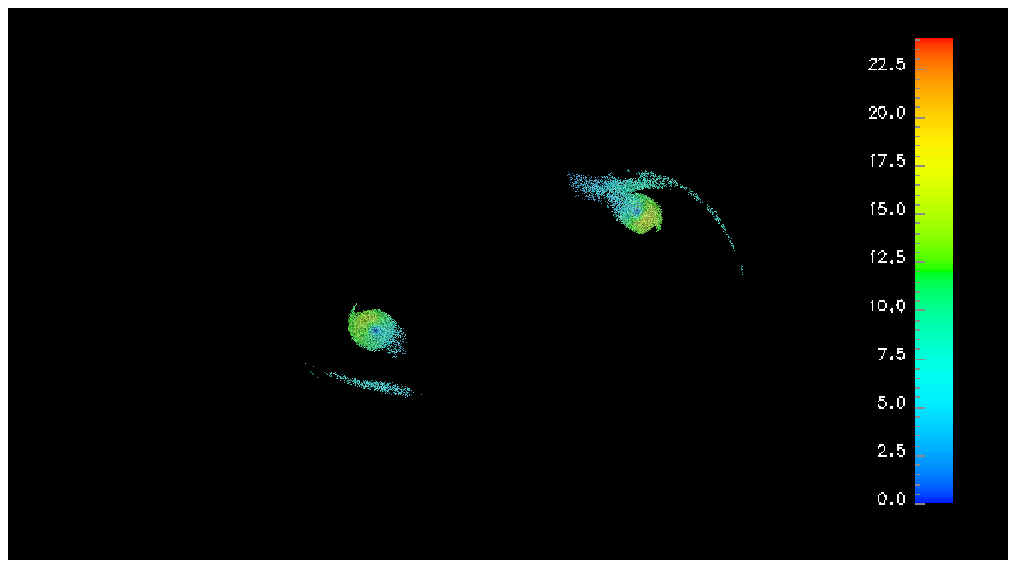} \\
\includegraphics[width=3.2in]{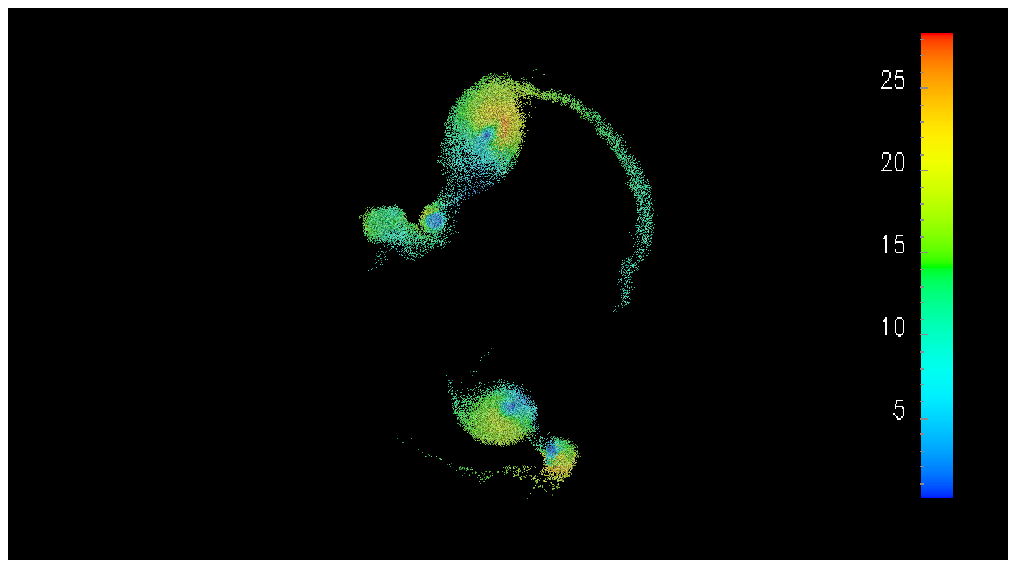} &
\includegraphics[width=3.2in]{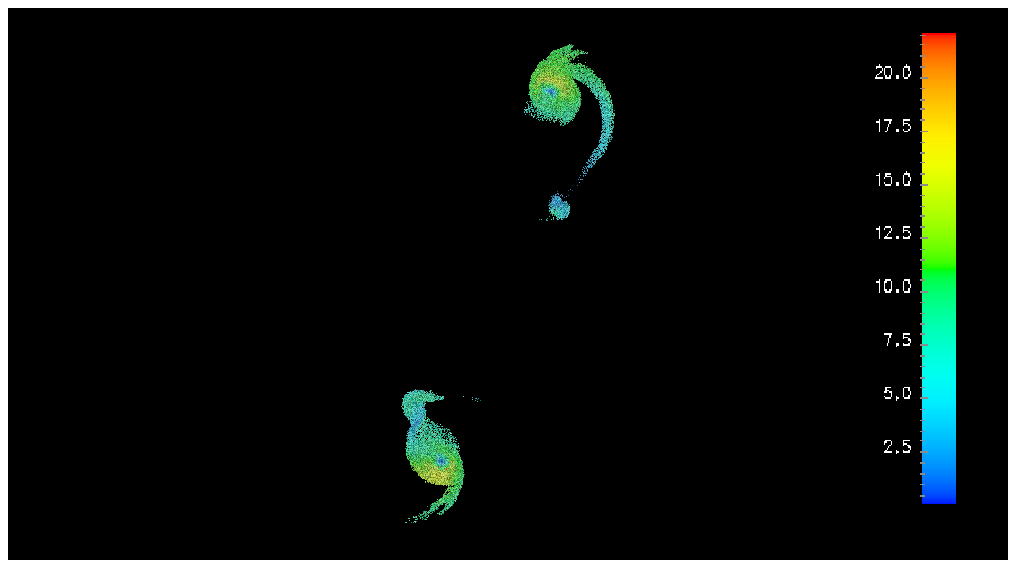}\\
\includegraphics[width=3.2in]{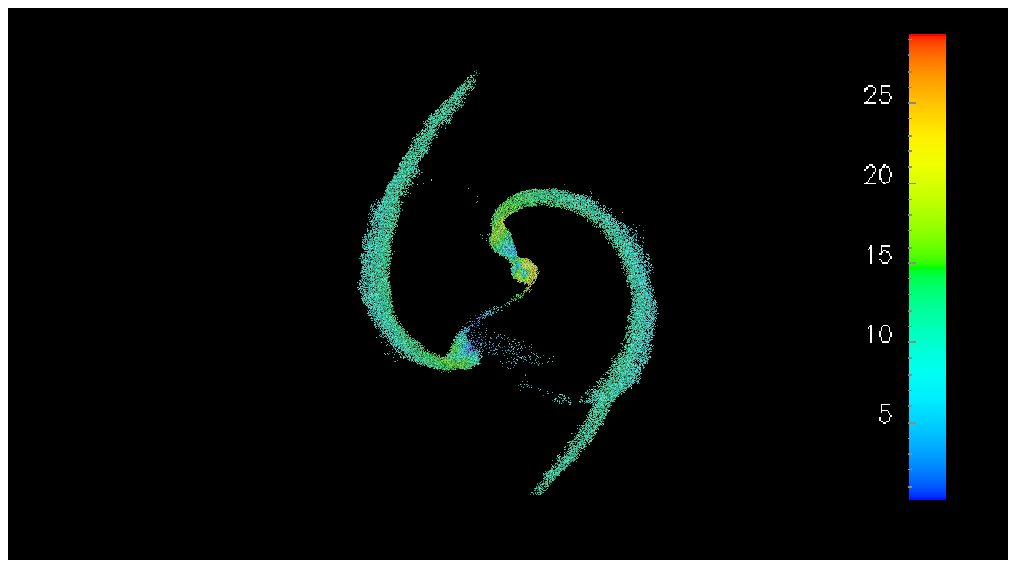} &
\includegraphics[width=3.2in]{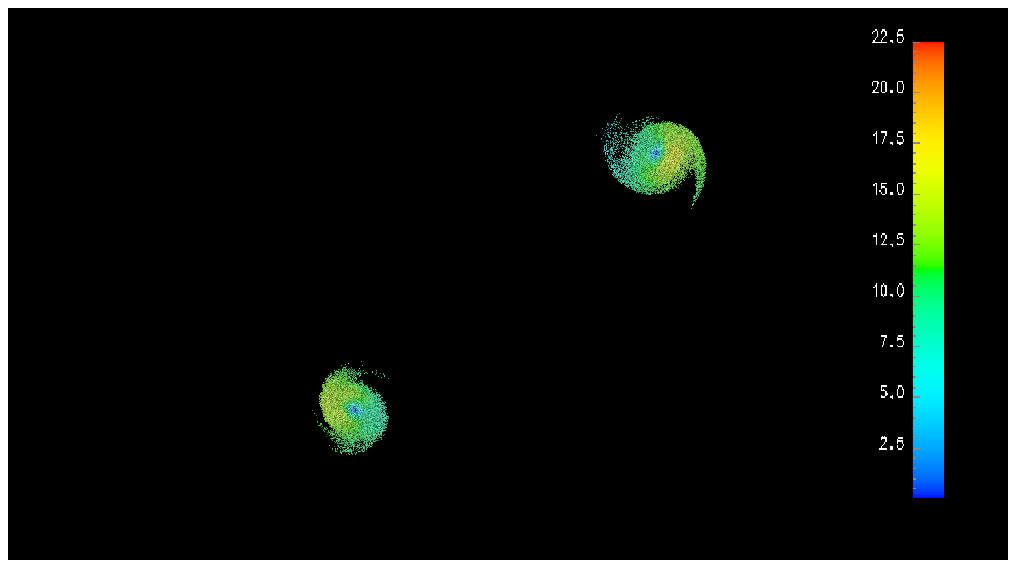} \\
\end{tabular}
\caption{\label{Vel3D} Iso-velocity 3D plot; each model is normalized with
its corresponding sound speed, ordered from top to bottom according
to the mass of the parent core:
m5b021 (first line right); m2p5b013 (second line left), m2p5b014
(second line right); m1p5b011 (third line left), m1p5b014 (third
line right); m1b045 (fourth line left), m1b014 (fourth line right)
and m0p75b045 (fifth line left), m075b014 (fifth line right).}
\end{center}
\end{figure}

Now we shall discuss the velocity scale of the particles
forming the fragments; see Fig.\ref{Vel3D}. These plots are 3D
representations formed by all the particles satisfying the following
selection criteria: (i) they are located within the central region
of the core, such that their projected radius
$r_{2d}=\sqrt{x^2+y^2}$ is $r_{2d}< r_{2dmax} \equiv 0.2\, R_0$,
irrespective of their $z$ coordinate; (ii) they have a density
higher or equal than the minimum density value given in advance by $
\log_{10} \left( \rho_{2dmin}/\rho_0 \right)=4.0$. These parameters
correspond to a minimum density $\rho_{2dmin}=3.82 \, \times \,
10^{-14}\,$g cm$^{-3}$ and a maximum radius of $r_{2dmax}=668$ AU.
This selection procedure is similar to the one we used earlier to
define a fragment. So, in this section, more
particles were considered to make the 3D plots: in the range
of 500-800 thousand, approximately.

We wish to point out that models have different sound speeds, $c_0$, which
are shown in column three of Table~\ref{tab:sep}. However,
of this, in Fig.\ref{Vel3D} we plot the magnitude of the velocity vector,
normalized with the sound speed, so that we will now use a
Mach $\equiv v/c_0$, as the unit of velocity to describe our results.

In all the models, we observed that: (i) only very few
particles reach very high velocities, marked with red color in the plots;
it should be noted that these ultra-fast particles are located
very close to the central region of the fragment and are velocity-isolated, as
no other neighboring particles have similar velocities;  (ii) particles
located exactly in the center of the fragment have the smallest
velocity, which can even range from 0.01 to 0.1 Mach, marked with blue
color in the plots; (iii) spiral arms are connected to their
fragments by particles with small
velocities, marked with soft blue or aqua color in
the plots; although not all of them are visible in these 3D plots, as
their density is probably not high enough to satisfy our density
selection criteria; (iv) the particles surrounding the innermost
region of the fragments have
intermediate velocities, marked with green color in the plots. These
are the already rained particles from the spiral arms on the fragment.
The particles that are still raining get the fragment through
contact regions between the spiral arms and the fragments, where the
raining particles have radial speeds it is marked with yellow
color in the plots; (v) the co-existence of different color scales in
a fragment indicates that there is a strong velocity gradient.


\section{Concluding Remarks}
\label{sec:conclu}

In this paper we have considered the gravitational collapse of a rotating
core using a spherical shell populated with SPH particles in order
to represent the core at the initial simulation time.

First, we observed that this mesh geometry appropriately represents
the relevant initial physics of the core, including the uniform
density distribution and the rigid body rotation. There is an excess
in density in early simulation time, as can be seen in
Fig.\ref{DenMaxP}, which is likely to be a consequence of the huge
surface density of the innermost radial shell. Fortunately, the
particles quickly adjust themselves and thus the core truly begins
its collapse some time later.

Second, the approximation strategy followed in this paper, that of
changing the core mass while keeping the core radius unchanged, can be
replaced by another, equally valid strategy, for instance, one in
which both the core mass and the radius are changed while the average
core density is unchanged. Hence, in this paper we can discuss
certain results obtained for a family of similar size cores
with slightly increasing total mass, while for the latter case one
could discuss about of a family of similar density cores with slight
mass and size variations.

Third, we observed that the more massive the initial core, the
lower its tendency to obtain the desired binary system formed via
the separation of the embryonic mass condensations. We thus prevented
their merging by providing more initial rotational energy to the core.
From the schematic configuration space reported in
Fig.\ref{Radr_Ejes}, we conclude that it is more likely to have a
binary system formed out of a small mass parent core and therefore
the mass of the binaries are expected to be small as well.

Fourth, in order to calculate the integral properties and the
velocity distributions of the obtained fragments, we took particles
by applying selection criteria based on two parameters $lrho_{min}$
and $r{max}$ whose values were fixed in advance. The selected
particles were those that had a density greater than or equal to $\rho_0\,
\times 10^{lrho_{min}}$ and a position radius $r<r_{max}$. Thus, one
would expect slight differences in the reported results as they are
definition-dependent.

Nevertheless, we find that there is a clear correlation between
the mass of the obtained fragments and the mass of the initial
collapsing core. In fact, the masses of the fragments are within the
observational range reported by \cite{obs}.

\section*{Acknowledgements}
GA. would like to thank ACARUS-UNISON for the use of their
computing facilities in the development of this manuscript.


\end{document}